# Optimal Future Sub-Transmission Volt-Var Planning Tool to Enable High PV Penetration

Quan Nguyen, Xinda Ke, Nader Samaan, Jesse Holzer, Marcelo Elizondo, Huifen Zhou, Zhangshuan Hou, Renke Huang, Mallikarjuna Vallem, Bharat Vyakaranam, Malini Ghosal, Yuri V. Makarov, *Member, IEEE*.

*Abstract* – This paper proposes a reactive power planning tool for sub-transmission systems to mitigate voltage violations and fluctuations caused by high photovoltaic (PV) penetration and intermittency with a minimum investment cost. The tool considers all existing volt-ampere reactive (var) assets in both sub-transmission and distribution systems to reduce the need of new equipment. The planning tool coordinates with an operational volt-var optimization tool to determine all scenarios with voltage violations and verify the planning results. The planning result of each scenario is the solution of a proposed optimal power-flow framework with efficient techniques to handle a high number of discrete variables. The final planning decision is obtained from the planning results of all selected violated scenarios by using two different approaches—direct combination of all single-step solutions and final investment decision based only on the scenarios that are representative for the power-flow voltage violations at most time steps. The final planning decision is verified using a realistic large-scale sub-transmission system and 5-minute PV and load data. The results show a significant voltage violation reduction with a less investment cost for additional var equipment compared to conventional approaches.

## I. NOMENCLATURE

**Sets**

| | |
|---|---|
| $E$ | Set of lines |
| $G_i$ | Set of generators at bus $i$ |
| $K_i$ | Set of blocks of switched shunt at bus $i$ |
| $L$ | Set of target load buses |
| $N$ | Set of buses |
| $S$ | Set of buses having switched shunts |

**Indices**

| | |
|---|---|
| $c$ | Circuit number |
| $i, j$ | System buses |
| $k$ | Switched shunt block |
| $m$ | Generator number |
| $t$ | Time step |

**Parameters**

| | |
|---|---|
| $b_i^s$ | Shunt susceptance at bus $i$, if any |
| $b^{\Delta_{ik}}$ | Step size of shunt block $k$ at bus $i$ |
| $b_{ijc}$ | Series susceptance of circuit $c$ between bus $i$ and $j$ |
| $b_{ijc}^C$ | Shunt susceptance of circuit $c$ between bus $i$ and $j$ |
| $d_i^P, d_i^Q$ | Real and reactive power demand at bus $i$ |
| $f$ | Objective function |
| $g_{ijc}$ | Series conductance of circuit $c$ between bus $i$ and $j$ |
| $g_i^s$ | Shunt conductance at bus $i$, if any |
| $I_{ijc}^R, I_{ijc}^I$ | Real and imaginary parts of the current in circuit $c$ between bus $i$ and $j$ |
| $\bar{I}_{ijcap}^2$ | Maximum current limit of the line between bus $i$ and $j$ |
| $P_i^s$ | PV real power generation at bus $i$ |
| $p_m^{sch}$ | Scheduled real power from generator $m$ |
| $P_i^{maxc}$ | Maximum solar active power curtailment for solar at bus $i$ |
| $Q_i^{DR\max}, Q_i^{DR\min}$ | Maximum and minimum limit of reactive power from demand response at bus $i$ |
| $Q_m^{\max}, Q_m^{\min}$ | Maximum and minimum reactive power limit of generator $m$ |
| $Q_i^{s\max}, Q_i^{s\min}$ | Maximum and minimum reactive power limit of aggregated solar at bus $i$ |
| $s_i^l$ | Target voltage at load bus $i$ |
| $S_i$ | Maximum power rating of the PV at bus $i$ |
| $x_{ik}^{\max}$ | The number of steps in shunt block $k$ at bus $i$ |
| $x_{ik}^*$ | Initial number of steps switched on of switched shunt block $k$ at bus $i$ |
| $v_m^{sch}$ | Scheduled voltage of PV bus $m$ |
| $V_i^{target}$ | Target voltage at load bus $i$ |
| $V_i^{db}$ | Dead band for target voltage at load bus $i$ |
| $\underline{V}_i, \bar{V}_i$ | Lower and upper voltage limits at bus $i$ |
| $\tau_{ijc}, \phi_{ijc}$ | Tap ratio and phase shift of the transformer between buses $i$ and $j$, if any |

**Variables**

| | |
|---|---|
| $P_i^c$ | Real power curtailment from PV at bus $i$ |
| $P_m^g$ | Real power from generator $m$ |
| $Q_i^{DR}$ | Reactive power from demand response at bus $i$ |
| $Q_i^s$ | Reactive power from PV at bus $i$ |
| $Q_m^g$ | Reactive power from generator $m$ |
| $V_i^R, V_i^I$ | Real and imaginary parts of the voltage at bus $i$ |
| $V_m^g$ | Voltage of generator $m$ |
| $x_{ik}$ | Number of steps switched on of switched shunt of block $k$ at bus $i$ |
| $X_i^+, X_i^-$ | Binary variables that represent the decision to install capacitors and inductors at bus $i$ |
| $B_i^+, B_i^-$ | Susceptance of the installed capacitor or inductor at bus $i$ |

## II. INTRODUCTION

Increasing distributed energy resources (DER) in modern power systems allows an intensive bi-directional interaction between distribution and sub-transmission sides [1]. High penetration of grid-edge, inverter-based photovoltaic (PV) might result in significant voltage fluctuations at both distribution and sub-transmission levels due to unavoidable PV intermittency. Such a voltage quality problem can be effectively mitigated by proper reactive power planning (RPP). The traditional objective of RPP is to regulate system voltage by selecting appropriate locations, types, and sizes of static and dynamic reactive power resources such as capacitor banks, static var compensators, and static compensators to achieve a desired voltage requirement with a minimum investment cost [2].

Traditionally, RPP greatly depends on the experience of power system planning engineers, which is apparently not reliable. The first analytical approach of RPP for voltage stability analysis is based on reactive power-to-voltage analysis [3][4]. However, this method is only suitable for initial screening instead of making the final planning decision due to its insufficient accuracy [5]. In [6], trajectory sensitivity is used to identify the locations and sizes for dynamic var support to mitigate the short-term voltage instability following a large disturbances. This approach, however, does not guarantee a minimum investment cost or system losses, which are common requirements in the planning stage.

An alternative RPP approach that has been adopted in most of the other existing works is based on optimal power flow (OPF) [7]. The objectives for these optimizations include the minimization of var investment cost, fuel cost, system losses, and voltage profile deviation for long-term analyses. The optimization problem characteristic, however, is mainly determined by the formulated constraints. OPF-based RPP is characterized by nonlinear power balance constraints at all buses and line flow limits as well as lower and upper bounds of the state and control variables, which can be both continuous and discrete [8]-[14]. Recently, dynamic behavior of the reactive power compensation has been incorporated in dynamic RPP to address short-term voltage stability following a contingency [6], [15]-[18]. The objective of these RPP studies is to identify locations of dynamic reactive power compensation to avoid slow voltage recoveries or voltage collapses. In addition to the constraints of RPP in normal operating conditions, these works incorporate additional constraints to achieve reliable operations under and after contingencies, which results in security-constrained OPF models. However, to the best of our knowledge, little attention has been paid to RPP considering renewable energy integration. References [17][18] propose a method for optimally locating and sizing dynamic reactive power compensation for large-scale integration of wind generation. On the other hand, PV has not been sufficiently considered as an asset to regulate voltage in RPP. With the increasing PV penetration in the grid, power system planning engineers need to consider smart inverter-based PV as a potential source of reactive power support in RPP. The coordination of var assets in both sub-transmission and distribution sides also needs to be considered to reduce the number and capacity of additional var resources [19][20].

All of these OPF and security-constrained OPF models are nonlinear, nonconvex optimization problems with discrete variables, which are hard to solve in polynomial time (NP-hard). Several approaches have been suggested in the literature to enhance the capability of solving such NP-hard problems. The linearization method described in [8][11][14] takes advantages of reliable and available mixed-integer linear programing solvers to obtain the global optimum. However, application of these approaches is limited due to the inaccurate linearized model compared to the original nonlinear model. Efforts to directly solve the nonconvex mixed-integer nonlinear programing (MINLP) are described in [8][15][18]. The disadvantage of these approaches is that they cannot guarantee global optimum. To achieve global optimum for the original nonconvex model, heuristic methods based on intelligent searches such as simulated annealing, evolutionary algorithms, and Tabu Search are studied [9][10][12][13]. However, these methods require significant computing time especially in large-scale systems. Therefore, the Bender decomposition method is applied in [21] to solve the RPP for large-scale systems, taking into account the discrete nature of var resources such as switched capacitors banks. However, this work only considers a few problematic scenarios, and it is not effective to RPP problems with high number of integer variables.

This paper proposes an Optimal Future Sub-Transmission Volt-Var Planning Tool (OFuST-VPT) at the sub-transmission level for long-term reactive planning to mitigate voltage fluctuation due to PV intermittency. The proposed planning tool ensures a minimum investment over multiyear, high-resolution PV and load data, taking into account both the existing var resources in sub-transmission system and PVs at the connected distribution systems. The novelty and uniqueness of the proposed planning tool are as follows:

- OFuST-VPT is developed based on a MINLP OPF model, so an effective relaxation approach is proposed to handle a huge number of integer variables that represent the candidate locations and discrete sizes of the potential var resources. A realistic large-scale sub-transmission system is used to demonstrate the capability and robustness of OFuST-VPT in determining additional var resources to achieve a desired voltage stabilization and system loss reduction.
- OFuST-VPT closely coordinates with a Coordinative Real-time Sub-Transmission Volt-Var Control Tool (CReST-VCT) [19]. CReST-VCT is an operational tool that aims to minimize total system losses and stabilizes regulate voltage by optimizing the existing var resources. CReST-VCT determines the scenarios with voltage violations as the input for OFuST-VPT. CReST-VCT is also used to verify the planning decision made by OFuST-VPT.
- Compared to the existing RPP works, the investment decision made by the proposed planning tool OFuST-VPT is based on multiple scenarios with different magnitudes and locations of voltage violations. Thus, another uniqueness of OFuST-VPT is two conservative and machine-learning based approaches to determine the final investment decision based on the planning results of all individual scenarios.

III. PROPOSED OPTIMAL FUTURE SUB-TRANSMISSION VOLT-VAR PLANNING TOOL (OFuST-VPT)

OFuST-VPT deploys the OPF approach to minimize the investment cost on var equipment at the sub-transmission level subjected to desired operational and planning constraints. As mentioned above, one crucial feature of OFuST-VPT compared to the existing approach is that var resources in distribution systems are considered in the planning stage, which is both compliant with the IEEE 1547 standard and beneficial for reducing investment cost. This section discusses the main features including the interface with the operational tool CReST-VCT, optimization model, and solution approach of OFuST-VPT.

A. *The Expected Roles of Smart-Inverter-Based PVs in Operation and Planning*

Recent developments in power electronics and control provide smart inverters advanced features to support grid voltage and frequency. With the rapid penetration of distributed PV generated power in distribution systems as a potential major participant of power system, these features could play an important role to strengthen the reliability and resiliency of the system. Therefore, the latest national DER standard IEEE 1547-2018 thoroughly describes specific requirements and guidance for the next generation of smart inverters to support power grid [22].

These new features of PV smart inverters not only affect the operation of power systems but also impact the traditional planning process in sub-transmission systems. Adept planning for sub-transmission systems must take into account the contribution of distributed smart-inverter-based PV power as valuable volt-var and volt-watt assets. Specifically, a voltage-load sensitivity matrix [23] was developed in the distribution management system to calculate the upper and lower limits of the real and reactive power from all distributed PVs and forecast the real power operating point in distribution systems for every 5 minutes. This data is provided to the planning process of OFuST-VPT incorporated in the transmission energy management system as given parameters. With the information about real and reactive power support from distribution systems, it is expected that the final investment cost for additional var devices from OFuST-VPT would be significantly less than the traditional approach that ignores the contribution of distributed PV power.

### B. Main Steps of OFuST-VPT

Fig. 1 shows the flow chart of OFuST-VPT. The bases for determining the investment in var equipment are the available annual load and PV data and the given cost for var equipment. For each year within the planning horizon, several load and PV deployment scenarios that produce extreme operating conditions such as voltage violations are selected. The OPF model, which is described in Section II.B, is solved to determine the optimal locations and sizes of var equipment and the corresponding investment cost for each scenario. The final planning decision for the studied year is made based on the OPF results of all scenarios.

The locations and sizes of var equipment of the final planning decision of each year are then incorporated into the existing network model to perform the operational tool CReST-VCT with a full year's load and PV data. The resulting performance of the network from a voltage management perspective with the additional var equipment is investigated. If the determined investment plan cannot help the system achieve desired voltage requirements, the planning optimization process is repeated. The scenarios with unacceptable voltage violations, which are based on both magnitude and duration, are considered as additional input scenarios for the subsequent planning process.

With the proposed scheme, the final RPP decision guarantees a satisfactory voltage requirement for the entire simulated year. The RPP process then progresses to the following year within the planning horizon with the incorporation of all var equipment installation from the previous years. Among all available var resources at the sub-transmission system, this work focuses on, but is not limited to, two options, which are capacitors and inductors.

### C. Selection of Extreme Scenarios and Candidate Locations for Additional var Resources.

To obtain the scenarios with voltage violations for OFuST-VPT in a year within the planning horizon, AC power flow is conducted for 1-year period with 5-minute data resolution of generators, PV, and loads. Based on the voltage magnitude $V_{i,t}$ at a target load bus $i$ from the power-flow solution at time step $t$, a voltage deviation index $V_t^{dev}$ for this time step is calculated as follows:

$$V_t^{dev} = \sum_{i \in L} |V_{i,t} - V_i^{target}|. \quad (1)$$

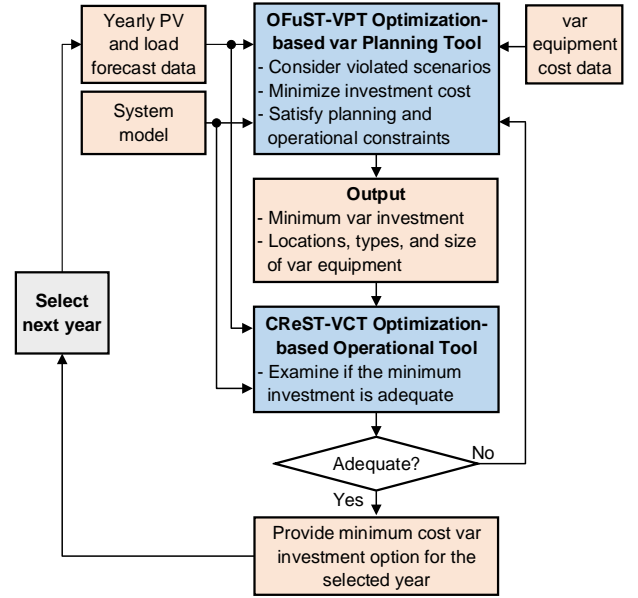

Figure 1: Flow chart of the proposed planning tool OFuST-VPT.

The voltage deviation indexes for all time steps in the entire simulated year are ranked in reducing order. A chosen number of time steps with the most severe voltage deviations are selected and used as the input scenarios for the OFuST-VPT in the particular year.

In OFuST-VPT, the candidate locations of additional var resources are chosen as the target load buses. Such a selection is reasonable since local var support reduces reactive power flow in the lines and thus transmission losses.

### D. Optimization Model of OFuST-VPT

The optimization formulation of OFuST-VPT corresponding to one scenario determined from Section III.C is written in rectangular coordinates. Let $N$, $E$, $G$, and $L$ be the set of buses, lines, generator buses, and target load buses respectively. In each line $(ij) \in E$, let $g_{ij}$, $b_{ij}$, and $g_{ij}^C$ denote its series conductance, series susceptance, and shunt susceptance, respectively. Let $S$ denote the set of buses having switched shunts and $K_i$ denote the set of blocks of the switched shunt at bus $i$. The set of candidate locations for installing additional var equipment is denoted by $C$.

The state variables $x$ consist of the real and imaginary parts $V_i^R$ and $V_i^I$ of bus voltages at all nodes of $N$. Because OFuST-VPT considers all existing var resources, the decision/control variables $u$ include reactive power $Q_i^g$ from generators at bus $i$, integer variable $x_{ik}$ representing the number of steps switched on at bus $i$, and real power curtailment $P_i^c$ as well as aggregated reactive power support $Q_i^s$ from the distributed PV and distributed capacitor banks in the interconnected distribution systems at bus $i$. The other components of $u$ are the planning variables $X_i^+$, $X_i^-$, $B_i^+$, and $B_i^-$, which represent the decision of installing an additional capacitor and inductor at bus $i$ as well as the corresponding susceptance.

The equality constraints are the real and reactive power balance equations at all buses, which are given as follows:

$$0 = \sum_{k \in G_i} P_k^g - d_i^P + (P_i^s - P_i^c) - [(V_i^R)^2 + (V_i^I)^2] g_i^s$$
$$- V_i^R (\sum_{ij \in E} I_{ij}^R + \sum_{ij \in E} I_{ij}^R) - V_i^I (\sum_{ij \in E} I_{ij}^I + \sum_{ij \in E} I_{ij}^I), \quad (2)$$

$$0 = \sum_{k \in G_i} Q_k^g - d_i^Q + Q_i^s + Q_i^{DR}$$
$$- V_i^R (\sum_{ij \in E} I_{ij}^I + \sum_{ij \in E} I_{ij}^I) - V_i^I (\sum_{ij \in E} I_{ij}^R + \sum_{ij \in E} I_{ij}^R)$$
$$+ \left( b_i^s + \sum_{k \in K_i} b^{\Delta_{ik}} x_{ik} + B_i^+ - B_i^- \right) \left[ (V_i^R)^2 + (V_i^I)^2 \right], \forall i \in N \quad (3)$$

where the real and imaginary parts of the line currents are calculated as follows:

$$I_{ij}^R = \frac{1}{\tau_{ij}^2}(g_{ij} V_i^R - (b_{ij} + \frac{b_{ij}^C}{2}) V_i^I)$$
$$- \frac{1}{\tau_{ij}}(g_{ij} V_j^R - b_{ij} V_j^I) \cos(\phi_{ij}) + \frac{1}{\tau_{ij}}(g_{ij} V_j^I + b_{ij} V_j^R) \sin(\phi_{ij}), \quad (4)$$

$$I_{ij}^I = \frac{1}{\tau_{ij}^2}(g_{ij} V_i^I + (b_{ij} + \frac{b_{ij}^C}{2}) V_i^R)$$
$$- \frac{1}{\tau_{ij}}(g_{ij} V_i^I + b_{ij} V_i^R) \cos(\phi_{ij}) - \frac{1}{\tau_{ij}}(g_{ij} V_i^R - b_{ij} V_i^I) \sin(\phi_{ij}), \quad (5)$$

The inequality constraints that comprise the operational limits are summarized as follows:

$$|\sqrt{(V_i^R)^2 + (V_i^I)^2} - V_i^{target}| \leq V_i^{db}, \forall i \in L \quad (6)$$
$$(I_{ij}^R)^2 + (I_{ij}^I)^2 \leq \bar{I}_{ijcap}^2, \forall (ij) \in E, \quad (7)$$
$$V_i^g = v_i^{sch}, \forall i \in G, \quad (8)$$
$$P_{ik}^g = p_{ik}^{sch}, \forall k \in G_i, \quad (9)$$
$$Q_{ki}^{min} \leq Q_{ki}^g \leq Q_{ki}^{max}, \forall k \in G_i, \quad (10)$$
$$0 \leq P_i^c \leq P_i^{maxc}, \quad (11)$$
$$\underline{Q}_i^s \leq Q_i^s \leq \bar{Q}_i^s, \quad (12)$$
$$\frac{(P_i^s - P_i^c)^2}{S_i^2} + \frac{(Q_i^s)^2}{k^2 S_i^2} \leq 1, \quad (13)$$
$$0 \leq x_{ik} \leq x_{ik}^{max}, \quad (14)$$

Constraints (6) represents the maximum deviation requirement of the voltage at the target load bus $i$ from its target value $V_i^{target}$. This constraint can be incorporated as a minimization objective, it is intentional in this work to formulate the voltage deviation as a separate constraint. With different values of $V_i^{db}$, different sets of locations and investment costs for additional var equipment are obtained. System operators can based on those results to make a final planning decision accordingly to achieve an acceptable voltage violation.

The inequality constraint (7) shows the upper $\bar{I}_{ijcap}$ of the line current between bus $i$ and $j$ to represent the thermal limit.

The constraints related to a PV bus $i$ are shown in (8)-(10). While the real power generation $P_{ik}^g$ is fixed at the scheduled values $p_{ik}^{sch}$ based on the day-ahead market bidding results, the reactive power generation $Q_{ik}^g$ can vary within a given range.

The constraints of power generation from solar are shown in (11)-(13), in which $S_i$ is the power capability of the associated inverter. Both the net real and reactive power support from solar are variable, with the former is implicitly expressed by the real power curtailment $P_i^c$ from the maximum power point $P_i^s$. The bounds for switched shunt elements, in which $x_{ik}^{max}$ is the number of steps in block $k$, are shown in (14).

The inequality that represent the limits of the planning variables are shown as follows:

$$X_i^+, X_i^- \in \{0,1\}, \forall i \in C, \quad (15)$$
$$B^+ \in \{0\} \cup Q^{B+}, \forall i \in C, \quad (16)$$
$$B^- \in \{0\} \cup Q^{B-}, \forall i \in C, \quad (17)$$

where $Q^{B+}$ and $Q^{B+}$ are the sets of susceptances from the additional capacitors and inductors, respectively. It is important to note that these planning variables are discrete. The solution technique to deal with these discrete variables is described in Section IV.A.

The purpose of minimizing the total investment cost for the additional var equipment is given as follows:

$$\min f = \sum_{i \in C} \left( C_0 (X_i^+ + X_i^-) + B_i^+ C_i^+ + B_i^- C_i^- \right), \quad (18)$$

where $C_0$ is a fixed installation cost while $C_i^+$ and $C_i^+$ are the cost coefficients that vary with the susceptances $B_i^+$ and $B_i^-$ of the additional capacitors and inductors, respectively, at bus $i$.

IV. SOLUTION APPROACH

A. Relaxation for discrete variables

The optimization model (1)-(19) is a MINLP problem, which is nonconvex and NP-hard. The solution approach for the operational tool CReST-VCT, which is described in detail in [24], needs further improvements to be successfully applied to the volt-var planning OFuST-VPT due to the introduction of planning variables. A high number of the candidate locations of var equipment, which results in a high number of discrete variables $X_i^+$, $X_i^-$, $B_i^+$, and $B_i^-$, aggravates the nonconvexity of the model and results in convergence issues. Therefore, the planning variables are relaxed to be continuous, and (15)-(17) are substituted by the following constraints:

$$0 \leq X_i^+, X_i^- \leq 1, \forall i \in C, \quad (19)$$
$$X_i^+ B^{+min} \leq B^+ \leq X_i^+ B^{+max}, \forall i \in C, \quad (20)$$
$$X_i^- B^{-min} \leq B^- \leq X_i^- B^{-max}, \forall i \in C, \quad (21)$$

where ($B^{+min}$, $B^{+max}$) and ($B^{-min}$, $B^{-max}$) are the minimum and maximum susceptances of the additional capacitors and inductors. Constraints (20) and (21) guarantee that the values of $B_i^+$ and $B_i^-$ at candidate location $i$ are feasible regardless the need to install var equipment at this bus.

The continuous values of $B_i^+$ and $B_i^-$ in the relaxed solution is rounded to the closest upper values. If the corresponding values of $B_i^+$ and $B_i^-$ are less than $B^{+min}$ and $B^{-min}$, respectively, a decision would be made to invest in var equipment.

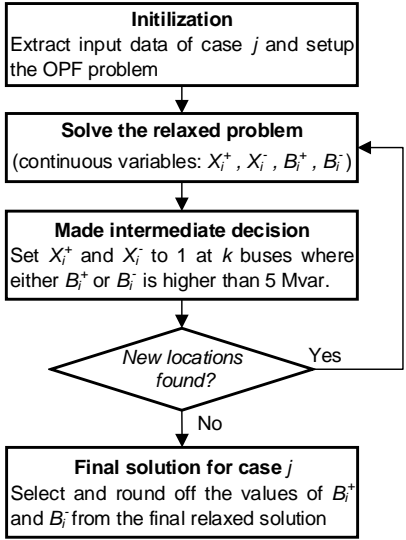

*Figure 2: Repetitive process to determine required locations and sizes of additional var resources in a scenario.*

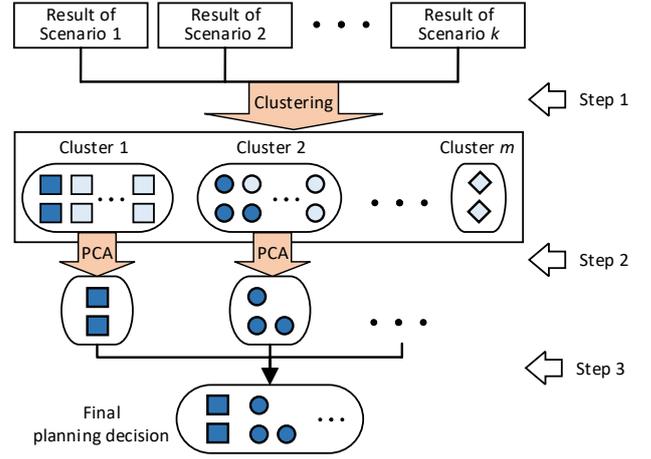

*Figure 3: Machine-learning based approach to make final planning decision from the planning results of separated scenarios.*

*B. Repetitive planning process for each studied scenario*

Because the discrete planning variables are relaxed to induce fast convergence, making an immediate final decision on the locations and sizes of the additional var resources from the obtained relaxed solution might lead to violations of system inequality constraints. Therefore, a repetitive approach as shown in Fig. 2 is implemented in OFuST-VPT to determine a conservative planning decision for each violated scenario.

First, the input data corresponding to Scenario $j$ is used to setup the relaxed OPF problem in Section II.D. The relaxed problem is solved by assuming planning variables are continuous. Based on the obtained relaxed solution, if the resulting value of either $B_i^+$ or $B_i^-$ is higher than the lower limit $B^{+\min}$ and $B^{-\min}$, respectively, the corresponding binary variable $X_i^+$ and $X_i^-$ is fixed to 1. The OPF problem is solved again. It is important to note that the susceptance sizes $B_i^+$ and $B_i^-$ are still considered as continuous variables. This repetitive process is terminated when no new locations of var resources are found. The final planning decision is made based on the relaxed solution at this stage. While the locations of var resources are already determined by the values of binary variables $X_i^+$ and $X_i^-$, their sizes are round off from the values of $B_i^+$ or $B_i^-$ in the last iteration.

*C. Two approaches to make final planning decision based on the solutions of multiple scenarios*

The solution approach mentioned above is applicable to a single scenario or time step solution of the power system. At different scenarios with different load and PV profiles, the voltage support requirement at the same area may change significantly. The different requirements on voltage support at different scenarios also result in different var resources investments. Therefore, two approaches are proposed to make a final investment decision on var resources based on the investment plans at different scenarios.

In the first approach, the final planning decision is the direct combination of all single-step solutions. This approach is conservative because the final investment incorporates the var requirements at all scenarios. For a candidate location $i$, the final size of the required reactive power resources is determined as follows:

$$B_i^+ = \max\{B_i^+(t)\}, \forall i \in C, \forall t \in T, \quad (22)$$
$$B_i^- = \max\{B_i^-(t)\}, \forall i \in C, \forall t \in T, \quad (23)$$

where $T$ is the set of problematic scenarios. Because this method is conservative, it results in a high investment cost.

Motivated from the economic disadvantage of the first approach, the main idea of the second method is to make final investment decision only based on scenarios that are representative for the power-flow voltage violations at most time steps. In other words, only scenarios that repeatedly require var resources at the same locations are considered in final investment plan. Therefore, the investment cost can be considerably reduced while guaranteeing sufficiently close performance compared to the first approach.

To achieve that goal, a method based on machine-learning as shown in Fig. 3 is proposed. First, a model-based clustering technique is applied to determine the optimal number of clusters based on the Bayesian Information Criterion [25]. The planning results from all scenarios are then divided into the chosen number of clusters. The results in a cluster are more similar to each other than to those in other clusters in terms of var resource locations and sizes. Based on the resulting clusters, the clusters with small numbers of components, such as Cluster $m$ in Fig. 3, are eliminated from the decision-making process.

Although all planning results corresponding to the scenarios in each cluster of the remaining clusters are extremely similar, they are not identical. Therefore, the investment cost can be further reduced by only considering the scenarios that are the most representative while ignoring the inferior ones in each cluster. This purpose can be achieved by applying the principal component analysis (PCA) to compute the contribution of each scenario to the most important principal components [26].

The final step of the second method is to combine the planning results from the most representative scenarios from the remaining clusters, which is based on (22) and (23).

Please note that both approaches are proposed to support system-planning engineers in making the final planning decision based on the available investment budget and corresponding system performance requirement.

TABLE I. TEST SYSTEM INFORMATION

| Parameters | Quantities |
|---|---|
| Buses | 3228 |
| Target load buses | 1360 |
| Generators | 177 units at 125 buses |
| PVs | 674 |
| Existing switched shunts | 366 |
| Candidate locations | 1360 |

## V. TEST SYSTEM AND RESULTS

This section shows the performance of the proposed OFuST-VPT framework, optimization model, and solution approach using a realistic sub-transmission system with a high number of candidate locations of var equipment. The formulated optimization model is solved in GAMS using KNITRO solver, and the solution feasibility is verified using PSS/E. The improvements of voltage profile using the OFuST-VPT planning solution, without and with PV support, are discussed.

### A. Test System Description

OFuST-VPT is applied to the Duke Energy Carolina (DEC) sub-transmission system to determine additional var resources to provide system voltage support. The description of DEC system is provided in Table I. Based on one power-flow case for the DEC system, we generated time series data at 1-hour resolution and corresponding power-flow scenarios for a full year. These time series data contain load profiles as well as PV generation at the buses where solar units are installed. PV data are leveraged from a PV integration study for the DEC system [27]. The main assumption for this study is 25% PV penetration by year 2025.

Based on the hourly power-flow results for the entire year, a voltage deviation metric for each scenario is calculated as in (6). The 21 scenarios with the highest voltage violations at the 44 kV and 100 kV levels are chosen as the input of OFuST-VPT. This planning plan with additional var resources is then applied to the DEC system for entire 1-year period to evaluate system performance in reducing voltage deviation.

### B. Simulation Results of a Single Scenario

One scenario in the 21 selected scenarios happens at 2:00 pm on April 4, in which voltage violations occurs at 86 load buses at the 44 kV level. A voltage violation tolerance of 0.002 phasor units (pu) and a step size of 5 Mvar for potential capacitors and inductors are chosen.

The relaxed planning solution from GAMS shows that additional inductors and capacitors are required at 92 and 14 locations, respectively. Fig. 4 shows the distribution of voltage violation at 1360 target load buses with the planning solution. The voltage violation at each bus is clearly less than the desired tolerance of 0.002 pu. Power flow with the additional inductors and capacitors is performed in PSS/E, which shows a similar result as the solution from GAMS. The fixed installation and Mvar costs for both capacitors and inductors are chosen to be $100,000 and $20,000/Mvar [28]. The resulting investment cost for this single scenario is $4.4 million.

Because most of reactive power support required from these 106 locations in the relaxed solution is small, only the equipment with ratings higher than 5 Mvar are selected, as shown in Fig. 2. The final planning solution for this studied scenario thus only includes eight locations to install capacitors and inductors with the sizes shown in Table II. By using the operation tool CReST-VCT with these additional var resources, the total voltage violation is reduced 6.12% compared to the base case when only the existing var assets are used.

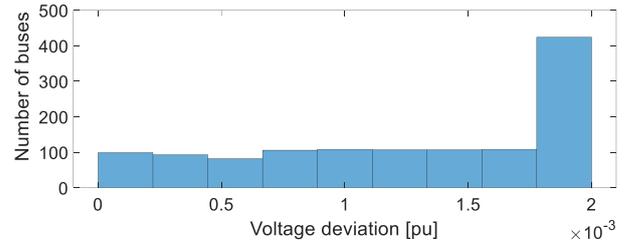

*Figure 4: Distribution of voltage violation at 1360 target load buses at 2:00 pm on April 4 with additional var support.*

TABLE II
REQUIRED VAR SUPPORT FROM CAPACITORS AND INDUCTORS

| Capacitors | | | Inductors | | |
|---|---|---|---|---|---|
| Bus | Relaxed Q [Mvar] | Final Q [Mvar] | Bus | Relaxed Q [Mvar] | Final Q [Mvar] |
| X1 | 21.37 | 20 | Y1 | 11.04 | 10 |
| X2 | 11.05 | 10 | Y2 | 9.78 | 10 |
| X3 | 9.56 | 10 | Y3 | 9.20 | 10 |
| | | | Y4 | 8.28 | 10 |
| | | | Y5 | 5.72 | 5 |

TABLE III
PLANNING RESULTS WITH AND WITHOUT VAR SUPPORT FROM PVS

| Number of locations | With Q support from PVs | | | | | |
|---|---|---|---|---|---|---|
| | Dead band of target load voltage (pu) | | | | | |
| | 0.002 | | 0.005 | | 0.01 | |
| | Relaxed | Final | Relaxed | Final | Relaxed | Final |
| Capacitors | 14 | 3 | 2 | 1 | 0 | 0 |
| Inductors | 92 | 5 | 17 | 0 | 3 | 0 |
| Cost ($M) | 15.4 | 2.9 | 2.4 | 0.3 | 0.34 | 0 |
| Number of locations | Without Q support from PVs (PF = 1) | | | | | |
| | Dead band of target load voltage (pu) | | | | | |
| | 0.002 | | 0.005 | | 0.01 | |
| | Relaxed | Final | Relaxed | Final | Relaxed | Final |
| Capacitors | 20 | 6 | 3 | 1 | 1 | 0 |
| Inductors | 208 | 25 | 89 | 9 | 31 | 2 |
| Cost ($M) | 36.1 | 11.1 | 13.5 | 3.1 | 4.4 | 0.6 |

Table III shows the required reactive power support and investment costs in the same scenario with and without reactive power support from the PV units located in the distribution feeders. In each case, different results are obtained with different tolerances of voltage violations. It is apparent from these results that a small tolerance of voltage violation results in higher investment cost for var equipment. More importantly, the required var support and investment cost are significantly less when distributed PV units are considered as var resources. This result thus shows the efficacy and generality of OFuST-VPT when including distributed PV units in the planning process.

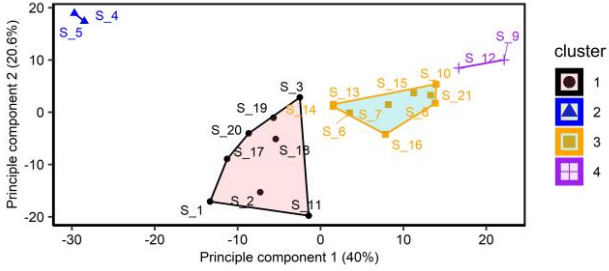

*Figure 5: Projection of the 21 selected scenarios onto the first two principal components of the PCA analysis, which account for 40% and 20.6% of the variance in the planning results of the 21 scenarios.*

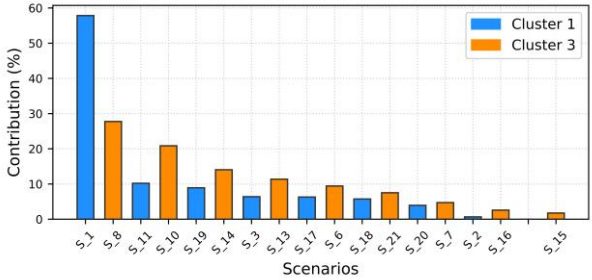

*Figure 6: Contributions of the scenarios in Clusters 1 and 3 to the first two principal components of each cluster.*

### C. Final Investment Decision Based on the Results from All Selected Scenarios

Apply the planning process in Section IV.B for all 21 selected scenarios results in 21 different planning results. As described in Section IV.C, two decision-making methods are used to make the final planning decision based on these 21 individual results.

The first method combines the planning results from all scenarios, which leads to an investment of $41.9 million with 114 additional capacitors and 22 additional inductors.

On the other hand, the second method first clusters the 21 scenarios into four groups using the model-based clustering method and corresponding Bayesian Information Criterion information, as described in Section IV.C. Fig. 5 shows the projection of the four clusters onto a two-dimension space, where the coordinates are the first two principal components obtained from a PCA analysis for all 21 scenarios. The numbers of scenarios in Clusters 1-4 are 8, 2, 9, 2, respectively. Only planning results corresponding to the scenarios in Clusters 1 and 3 are selected as the input for the subsequent PCA analysis as they have significantly more components than the other two clusters.

Fig. 6 shows the contribution of each scenario in Clusters 1 and 3 on explaining the variances retained by the first two principal components, which is obtained from the PCA analysis for each cluster. The planning result corresponding to Scenarios 1 and 8 are the most representative result in Cluster 1 and 3, respectively. In this work, only the planning results corresponding to these scenarios are chosen to make the final planning decision. By combining these two results, the final planning decision is made with an investment cost of $32.0 million with 86 additional capacitors and 17 additional inductors.

### D. Analysis on System Performance with the Final Planning Result from OFuST-VPT

As shown in Fig. 1, the capacitors and inductors determined from the final OFuST-VPT planning decision is added to CReST-VCT as additional installed var resources for the solving the OPF problems in several scenarios with voltage violations, including those outside of the aforementioned 21 scenarios. The total voltage violation in each scenario is calculated using (1) and a chosen dead band of 0.005 pu.

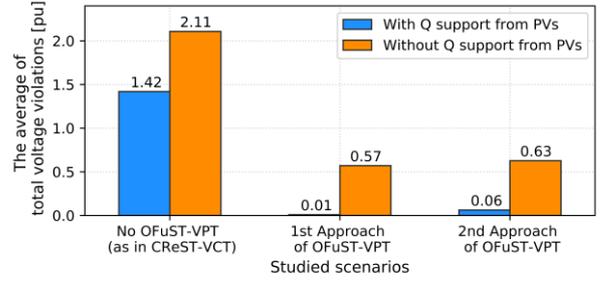

*Figure 7: Average of total voltage violations in the selected scenarios with and without using the OFuST-VPT solution and reactive power support from PV units.*

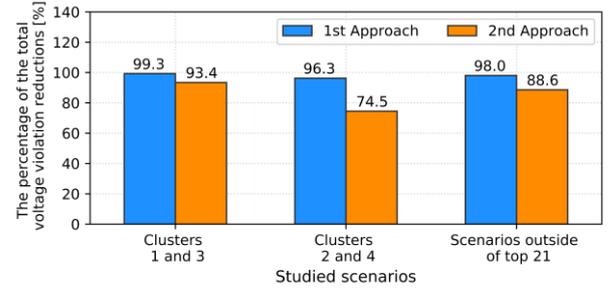

*Figure 8: Average reduction of total voltage violations compared to the base case (CReST-VCT) in three categories of scenarios using two investment approaches of OFuST-VPT.*

Fig. 7 shows the average of total load voltage violations in the selected scenarios with and without using OFuST-VPT solution and reactive power support from PV units. Regardless other var resources, reactive power support from PVs is shown to reduce an average of 0.56 to 0.69 pu of total voltage violation at 1360 load buses. On the other hand, the additional capacitors and inductors from the final OFuST-VPT planning decision contributes a significant 1.36 to 1.54 pu average voltage violation reduction, compared to that when considering only the existing var resources as in CReST-VCT. In addition, the voltage violation is almost eliminated when employing the reactive power support from PV units and the additional planning equipment.

Between the two approaches used to make the final planning decision in OFuST-VPT, the average total voltage violation with the additional var resources determined by the second approach is 0.05 pu higher than that when using the first approach. However, this voltage violation increase is marginal, given the approximated $10 million (25%) reduction in the investment cost of the second approach, as described in Section V.C.

As shown in Figs. 5–6, the final investment plan from the second approach is selected only based on scenarios that are representative for the power-flow voltage violations at Clusters 1 and 3. However, not just for the scenarios within Clusters 1 and 3, the target of the installed capacitor and inductor investment is to minimize voltage violations at all time steps on all target load buses. Therefore, in Fig. 8, we compare the performance of voltage violation reduction on all target load buses for scenarios at different clusters with the first and second approaches in OFuST-VPT. As can be seen from Fig. 8, both approaches can clearly reduce most of the voltage violations at target load buses for all scenarios. Even for those scenarios outside of the top 21 selected, the OFuST-VPT can still reduce 98% and 88.6% voltage violations on average. For scenarios within Clusters 1 and 3, the extra installed capacitors and inductors from the first and second approaches can significantly reduce 99% and 93.4%, respectively, of voltage violations for the original base case. The significant voltage violation reduction is because the scenarios at Clusters 1 and 3 are considered in the final investment plan from the first and second approaches. The first and second approaches can reduce 96.3% and 74.5% voltage violations on scenarios 2 and 4 on average. Compared with other two cases, the relatively low voltage violation reduction for the second approach for Cluster 2 and 4 is because the final investment plan did not consider any scenarios within Clusters 2 and 4.

## VI. Conclusion

This paper proposes OFuST-VPT as a reliable RPP tool in sub-transmission systems. The tool deploys an OPF-based framework that minimize the investment for var equipment while mitigating voltage violations and fluctuations caused by high PV penetration and intermittency. OFuST-VPT is verified using a realistic sub-transmission system and 21 scenarios with the highest voltage violations. Two approaches are proposed to make the final planning decision based on the solutions of multiple scenarios. The results show a significant voltage violation reduction with a reasonable investment cost. Therefore, OFuST-VPT will help power system planners to optimally add var resources in coordination with other existing resources such as generators, switched shunts, and PV inverters, especially when the coordinated operation of the existing assets in both the sub-transmission and distribution sides is not sufficient to increase PV penetration. This framework can also be extended for the planning of other resources such as energy storage and inverter-based PV units.

## Acknowledgment

This study is funded by the U.S. Department of Energy (DOE) SunShot Initiative as part of SunShot National Laboratory Multiyear Partnership (SuNLaMP). The project team wants to especially thank Mr. Jeremiah Miller, Dr. Guohui Yuan, and Dr. Kemal Celik, from the Systems Integration Subprogram at DOE's SunShot Initiative for their continuing support, help, and guidance.